# Generation of water-in-oil and oil-in-water microdroplets in polyester-toner microfluidic devices

Evandro Piccin [a]*, Davide Ferraro [b], Paolo Sartori [b], Enrico Chiarello [b], Matteo Pierno [b], and Giampaolo Mistura [b]**

[a]Departamento de Química, Universidade Federal de Minas Gerais, Belo Horizonte, MG, Brazil.
[b]CNISM and Dipartimento di Fisica e Astronomia G. Galilei, Università degli Studi di Padova, Padova, Italy.

*Corresponding authors at:*
* Departamento de Química, Universidade Federal de Minas Gerais, Av. Antônio Carlos 6627, Pampulha, Belo Horizonte, MG, Brazil. Tel.: +55 31 34096389. E-mail: evandrop@ufmg.br.
** Dipartimento di Fisica e Astronomia G. Galilei, Università degli Studi di Padova, Via F. Marzolo 8, 35131 Padova, Italy. Tel.: +39 049 827 7020. E-mail: giampaolo.mistura@unipd.it.

This paper demonstrates that disposable polyester-toner microfluidic devices are suitable to produce either water-in-oil (W/O) or oil-in-water (O/W) droplets without using any surface treatment of the microchannels walls. Highly monodisperse W/O and O/W emulsions were generated in T-junction microdevices by simply adding appropriate surfactants to the continuous phase. The dispersion in size of droplets generated at frequencies up to 500 Hz was always less than about 2% over time intervals of a couple of hours.

.

## 1. Introduction

Droplet-based microfluidics is an active research field that presents great potential for high-throughput chemical and biological analysis, synthesis of advanced materials, sample pretreatment, protein crystallization and encapsulation of cells [1-4]. In these devices, droplets of a fluid forming the so-called dispersed phase are carried by the stream of a second immiscible fluid identified as the continuous phase. Depending on the application, two major types of droplets are usually produced: water-in-oil (W/O) or oil-in-water (O/W) droplets.

The generation of stable droplets depends, among other things, on the wettability of the channel walls. The continuous phase should completely wet the channel walls, while the degree of wetting by the dispersed phase should be very low. Oils preferentially wet hydrophobic surfaces, whereas aqueous solutions preferentially wet hydrophilic surfaces. Thus, W/O droplets are usually produced in hydrophobic microchannels and O/W droplets are commonly generated in hydrophilic devices. Most published studies involve the production of W/O droplets because the majority of biological and chemical reactions occur in an aqueous solution [5-8]. This significantly simplifies the fabrication of microfluidic devices because it requires hydrophobic walls, a feature guaranteed by the use of common polymers. These include polydimethylsiloxane (PDMS) [9,10] and poly(methylmethacrylate) (PMMA) [11,12]. However, when O/W droplets are needed, the polymeric channel walls must be turned hydrophilic. Over the past few years various methodologies [6,7,13-15] have been devised to accomplish this task, which are usually technically demanding and time consuming. An alternative way consists in the addition of suitable surfactants to the continuous phase [5,16].

The direct-printing technology for fabricating toner-based microfluidic devices was first proposed by do Lago et al. in 2003 [17]. In this technique, the device layout is laser-printed on two sheets of polyester film that are then sealed together by thermal lamination for creating the channels. The toner printing acts as a glue in the lamination process. The depth of the channels (~ 12 μm) was determined by the height of the two toner layers that are laminated together. More recently, this fabrication process was modified in order to produce multilayered polyester-toner devices with larger channel depths [18]. In this latest methodology, the microchannels are made by cutting a polyester sheet containing uniform layers of toner. This cut-through sheet is sandwiched between uncoated polyester sheets containing precut access holes. Thus, the depth of the channels is determined by the number and the thickness of the middle cut-through sheets and the toner layers. When compared to other rapid prototyping techniques, the toner-based technology presents some advantages owing to its very short fabrication time, independence of clean room facilities and low cost (~ $ 0.15 per device) [19]. Polyester-toner (PeT) devices have found widespread applications in many areas including clinical bioassays [19], microchip electrophoresis [20], DNA extraction and PCR amplification [18], and DNA analysis [21]. However, all of these applications were developed for conventional continuous flow systems and, considering the increasing demand for droplets based devices [1,2], it would be important to merge these two technologies.

In the following, we will present T-junctions made in PeT where the two immiscible liquids flow through two orthogonal channels and form droplets when they meet [22] (see Fig. 1). The droplets generated in this way are highly monodisperse and are separated by an equal spacing [1-4]. It will be demonstrated that these disposable PeT devices can be used for generating W/O and O/W droplets without any surface modification of the microchannels. Stable W/O droplets were obtained using hexadecane or mineral oil as continuous phases with the addition of Span 80, while O/W droplets were generated using aqueous solution with Triton X-100 as continuous phases and hexadecane or mineral oil as dispersed phases. Surfactants were added to the continuous phase to lower the interfacial energy, that is to facilitate the formation of new interfaces and to stabilize the formed emulsion droplets from coalescence during their motion [4,16].

## 2. Materials and methods
### 2.1 Chemicals

All chemicals used were of analytical reagent grade quality and were employed as received. Hexadecane (CAS number: 544-76-3, viscosity ~3 cP, density ~0.77 g/cc ), mineral oil (light) (CAS number: 8042-47-5, viscosity ~16 cP, density ~0.84 g/cc ), Span® 80, and Triton™ X-100 were purchased from Sigma-Aldrich. Polyester sheets (transparency films, model CG3300) were acquired from 3M. Distilled water was used in all experiments.

### 2.2 Fabrication procedure

The PeT microfluidic devices were fabricated by direct-printing combined with xurography [17, 23-25]. The fabrication process was similar to previously reported methodologies [18,19]. Polyester sheets (A4 size, 100-μm-thick) were first printed twice on both sides using a laser printer (Hewlett-Packard model 4250) with 600-DPI resolution and equipped with a black toner cartridge (model Q5942A). This process yields two layers of toner printing on each side of the polyester film. The T-junction layout for the devices was drawn using the AutoCAD 2011 software and the microchannels were created in small pieces (55 × 35 mm) of printed polyester (see Fig. 1) using a knife plotter (Craft Robo CC200-20, Graphtec). The cut-through printed polyester film was placed between two unprinted polyester layers; with the upper one provided with access holes made with a paper punch. The three polyester sheets were laminated using an office laminator at 150 °C. Finally, reservoirs were constructed by gluing small pieces of rubber tubes (4 mm of i.d. and 10-mm-long) with epoxy resin. Using this procedure, we were able to produce channels having a minimum width of 60 μm and heights of 120 μm. The width was set by the resolution of the knife plotter, while the height (measured by imaging the channel cross-section with an optical microscope) was set by the thickness of the middle polyester sheet (~100 μm) and the toner layers. The thickness of one layer of toner is about 7 μm as determined with a micrometer. Thus, the four toner layers result in a total height of about 28 μm, which decreases to about 20 μm after the lamination process. Photos of the final device and of channels cross-sections of different width are shown in the bottom part of Figure 1. For the smallest channels, the observed irregularities are also due to the cutting of the chip with a razor blade to open up the cross-section. More generally, the surface finish depends on the quality of the cutting plotter [25].

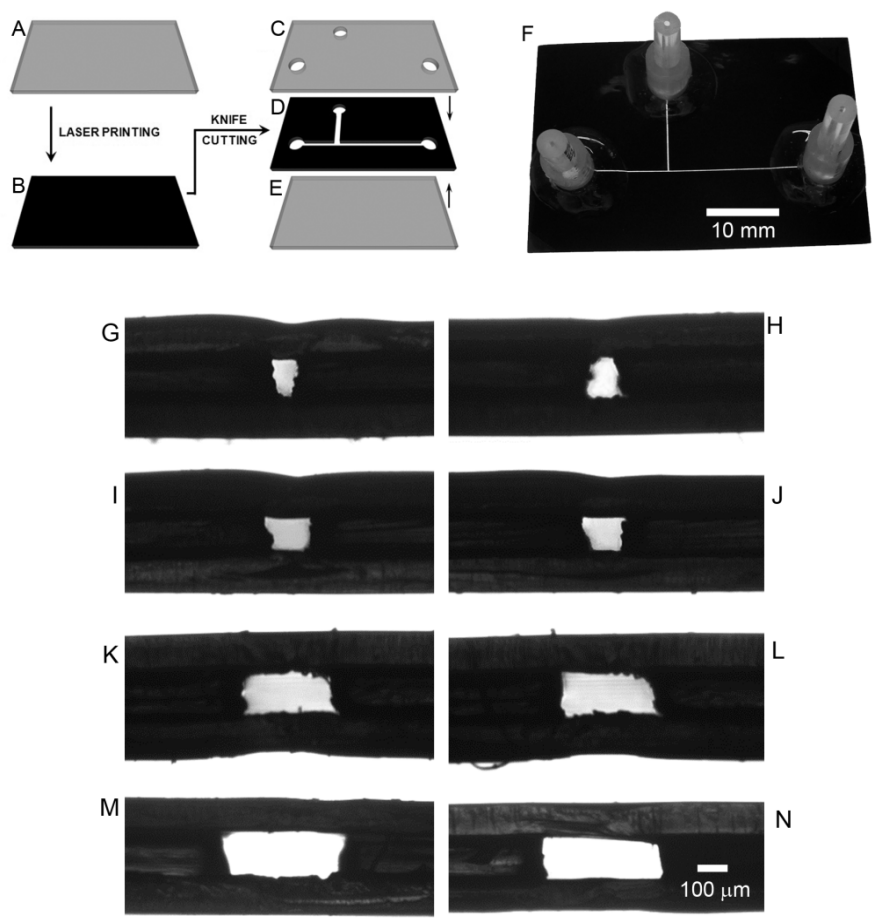

**Fig. 1.** Schematic diagram showing the main steps of the fabrication process: A) polyester film, B) polyester film toner-printed on both sides, C) polyester film with perforated access holes, D) microchannels cut on the printed polyester, E) polyester film used for sealing. F) Full view photograph of the T-junction microfluidic device constructed in PeT and utilized for the generation of W/O and O/W droplets. G)- N) Photographs of the cross-section of microchannels of different size. Each image refers to a different PeT chip. Characteristic widths: G)-H) 60 μm; I)-J) 90 μm; K)-L) 250 μm; M)-N) 350 μm.

**2.3 Droplets formation**

The microfluidic devices utilized for droplets generation were made with a T-junction layout [22] and had an overall size of 40 by 20 mm. All experiments were conducted in microfluidic channels with widths of ~250 μm and height of ~120 μm. We have used these devices continuously up to a couple of weeks without noticing any variation in their performance.

The liquids were introduced into the microfluidic devices through flexible polyethylene tubing (0.5 mm i.d.) and using two independent syringe pumps (PHD 2000, Harvard Apparatus) working at constant flow rates between 0.5 and 20 μL min$^{-1}$. The images of on-chip droplet formation were obtained using a monochrome camera (MV D1024 CMOS, Photonfocus) coupled to an inverted microscope (Eclipse Ti-E, Nikon). All data presented in the next section refer to as made chips. We have never reused a chip flowing different liquids in the channels.

**3. Results and discussion**

The static contact angle (CA) of water on the transparency (polyester) surface was measured to be 49 ± 2°, and this value compares well with recently reported data [26]. The partial hydrophilicity of the polyester surface together with the addition of suitable surfactants to the corresponding continuous phase were the key ingredients for the effective production of stable W/O and O/W droplets in the same PeT device. Since the wettability of thiolenic resins is similar to that of polyester, we expect that W/O and O/W droplets generators can also be successfully fabricated in thiolene by rapid prototyping techniques [27].

The first studies aimed at producing W/O droplets. Hexadecane and mineral oil (viscosity ~3 cP, density ~0.77 g/cc ) were evaluated as continuous phases in the absence and with different concentrations of the lipophilic nonionic surfactant Span 80. A series of pictures showing the formation of water-in-hexadecane and water-in-mineral oil droplets in the PeT T-junction is displayed in Fig. 2 for different surfactants concentrations (see also Movie 1 in Supplemental Material). In the absence of the surfactant, no droplets were obtained and a continuous water stream was formed parallel to the oil continuous phase (see Fig. 2A and 2B). Stable water-in-hexadecane droplets were obtained after adding 0.5% (w/w) of Span 80 to the continuous phase (Fig. 2C). However, 0.5% of Span 80 was not enough to generate water-in-mineral oil droplets (Fig. 2D). Stable water-in-mineral oil droplets were obtained only when the concentration of Span 80 was 1.0 % (w/w) (Fig. 2F).

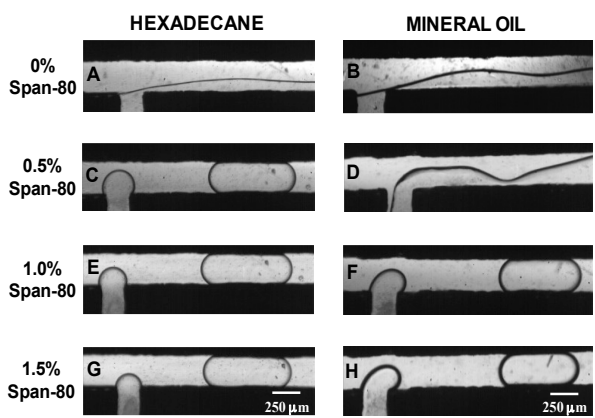

**Fig. 2.** Optical microscope images showing the generation of W/O droplets in the PeT devices. Hexadecane and mineral oil were used as continuous phases in the absence (A,B) and with increasing concentrations of Span 80 (C-H). The flow rate of the continuous and dispersed phases was 5 and 3 μL/min, respectively.

Hexadecane completely wet the polyester surface while pure mineral oil formed a CA=$10 \pm 2^{o}$, which remained practically constant ($9 \pm 2^{o}$) after the addition of 1.0 % (w/w) of Span 80, suggesting that the wettability of the channel walls by the oil phase was not affected by the presence of Span 80. Thus, the surfactant is facilitating the on-chip droplets generation by reducing the interface tension between the continuous and discrete phases and by preventing the coalescence of the resulting droplets [15]. The droplet length was found to be independent of surfactant concentration for both water-in-hexadecane (Fig. 2C, E, and G) and water-in-mineral oil (Fig. 2F and 2H) droplets. This behavior is expected once the minimum interfacial tension between the two liquids is achieved and this condition occurs whenever the surfactant concentration is over the critical micelle concentration (0.03% w/w and 0.025% w/v for hexadecane and mineral oil [5, 28]).

The production of O/W droplets in the polyester-toner devices was carried out using hexadecane and mineral oil as dispersed phases. Pure water and aqueous solutions containing different amounts of the surfactant Triton X-100 were tested as continuous phase. Fig. 3 displays optical microscope images about the formation of oil-in-water droplets (see also Movie 2 in Supplemental Material).

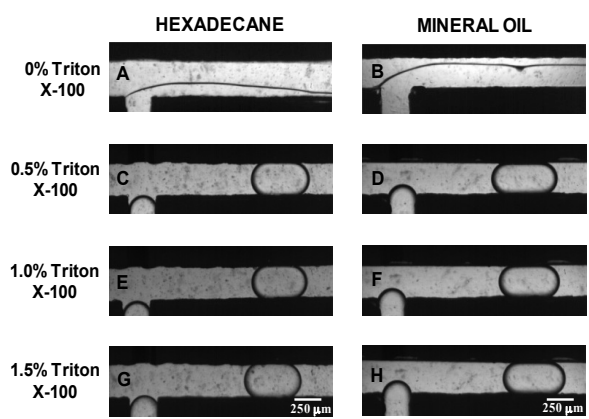

**Fig. 3.** Optical microscope images showing the generation of O/W droplets in the PeT microfluidic devices. Hexadecane and mineral oil were tested as dispersed phases, while pure water (A,B) and aqueous solutions containing increasing amounts of Triton X-100 (C-H) were used as continuous phases. The flow rate of the continuous and dispersed phases was 5 and 1 µL/min, respectively.

Again, without the surfactant, no droplets were formed and the two liquids formed two parallel streams (see Fig. 3A and 3B). The addition of 0.5% of Triton X-100 to water lead to the formation of reproducible and well defined hexadecane-in-water and mineral oil-in-water droplets (Fig. 3C and 3D). The reason is likely the enhanced wettability of the channel walls by the continuous aqueous solution. Actually, after the addition of 0.5% of Triton X-100, the CA of water was found to decrease from $49 \pm 2°$ to $13 \pm 2°$. As observed in the production of W/O droplets, the length of the O/W droplets also remained constant upon increasing the surfactant concentration to 1.0 and 1.5% v/v (Fig. 3E, F, G and H). As the critical micelle concentration of Triton X-100 in water is 0.015 % w/v [1], the interfacial tension between both phases is at its minimum value for all studied surfactant concentrations.

For biological applications, it is attractive to use a fluorocarbon oil as the continuous phase, as these oils are both hydrophobic and lipophobic, hence they have low solubility for the biological reagents of the aqueous phase and are well suited for inhibiting molecular diffusion between drops [9]. We have tested the compatibility of two common fluorocarbon oils, FC40 and HFE 7500, with Pe foils. They both completely wet the Pe surface and no indication of swelling was found after the oils were in contact with Pe for one day. Since their viscosities ( FC40 ~4.1 cP , HFE 7500 ~ 1.2 cP ) are quite similar to that of hexadecane, we expect that they can also be successfully used in PeT droplet generators.

For a more complete evaluation of the performance of these PeT chips, the production of droplets was studied over a time scale of about 2 hours. Sequences of more than 50 droplets were acquired in time intervals of 60 seconds separated by 20 or 30 minutes. The length of the droplets was analyzed with a custom made program [29]. The graph of Fig. 4 displays the mean values $l$ of each sequence together with the accompanying dispersion $\sigma$ for the four different combinations tested in this work. It clearly appears that the stability in time of the droplet generators was quite good and this implies that the mechanical stability of the device, as well as that of the channels walls wettability, is quite good. For instance, after storing some used chips in a cabinet for a couple of months, we repeated the measurements with the same combination of liquids without finding any difference with their initial performance. In the case of PDMS channels which have been functionalized to become hydrophilic, this effect usually disappears after they have been exposed to air for a couple of hours [6].

---

[1] Sigma-Aldrich Co (1999) TX100, T9284 product information sheet (Sigma-Aldrich, St. Louis)

The four accompanying histograms in Figure 4 indicate the droplet size distributions corresponding to the empty symbols in the graph taken after 2 hours. The resulting polydispersity, i.e. the ratio σ/ *l*, was always less than 2%. The variations in *l* for the four combinations were due to the different flow rates employed in the measurements.

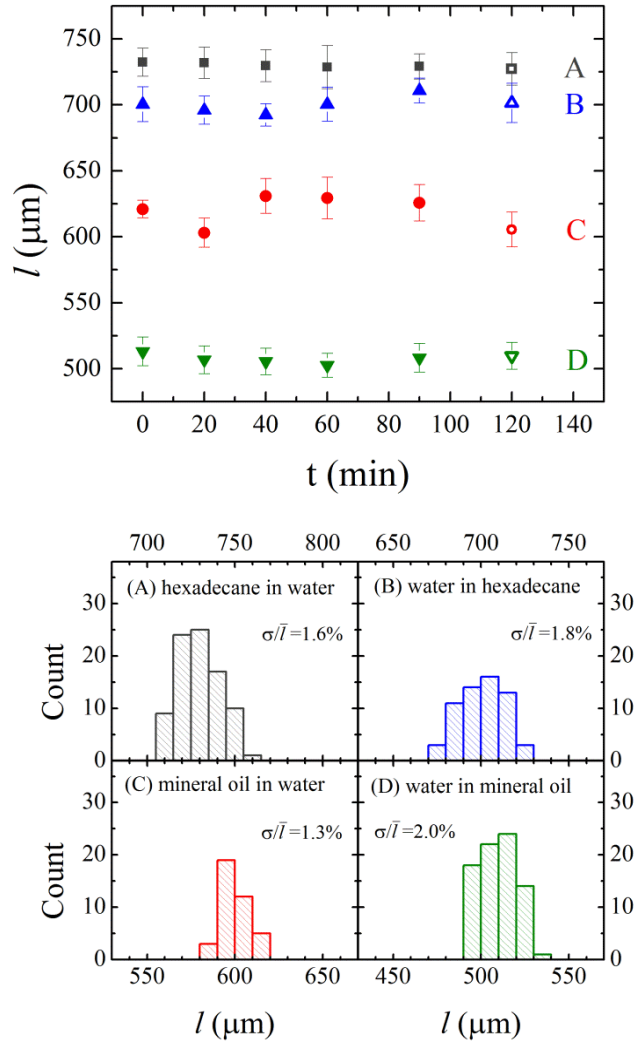

**Fig. 4.** Top: Stability over time of the length of droplets of different types produced with PeT devices. A datapoint corresponds to an average of at least 50 measurements. Each type of droplet has been produced in a different chip. (A) hexadecane-in-water, (B) water-in-hexadecane, (C) mineral oil-in-water, and (D) water-in-mineral oil. Bottom: Histograms showing the size distribution for the four types of droplets.

The data on the stability over time of the PeT droplet generators presented in Fig. 4 were acquired under the conditions summarized in Table 1.

**Table 1.** Fluidic conditions of the trains of droplets presented in Fig. 4.

| Dispersed phase | Continuous phase | Surfactant in the continuous phase | $Q_d$ (μL/min) | $Q_c$ (μL/min) | $Q_d/Q_c$ |
| --- | --- | --- | --- | --- | --- |
| Water | Mineral oil | 1.5% (w/w) Span® 80 | 2 | 5 | 0.4 |
| Water | Hexadecane | 0.5% (w/w) Span® 80 | 2 | 7 | 0.3 |
| Hexadecane | Water | 1.0% (v/v) Triton™ X-100 | 5 | 8 | 0.6 |
| Mineral Oil | Water | 1.5% (v/v) Triton™ X-100 | 2 | 10 | 0.2 |

We have also systematically analyzed the production of droplets as a function of the two inflows. Figure 5 shows the summary of these measurements. The graph displays the variation of the droplet length $l$ normalized to the channel width $w$ with the ratio of the flow rates of the dispersed $Q_d$ and continuous $Q_c$ phases. Each data point refers to a different chip and different combinations of the two liquids plus surfactants. The chips had the same nominal width w=250 μm and height h=120 μm. The vertical error bars are mainly due to the cross-section irregularities as shown in Figure 1. Based on this optical investigation, we estimated a deviation of the size w of about 50 μm. Despite the ample variations of all system variables, the data are found to collapse in a narrow region of the graph and the points are statistically consistent within one standard deviation. We believe that the observed scatter can be significantly reduced using a better cutting plotter. In the squeezing regime, the dynamics of break-up is dominated by the pressure drop across the droplet as it forms [30]. The length of the drops can be conveniently expressed with the following scaling equation

$$\frac{l}{w} = \alpha + \beta \frac{Q_d}{Q_c}$$

where α and β are two fitting parameters, whose value is mainly affected by the geometry of the T-junction [31]. The straight line appearing in the graph is the linear fit to all the data with α=1.8 and β=2.3. These values compare well with those in the literature [31] and confirm the robustness of PeT droplet generators.

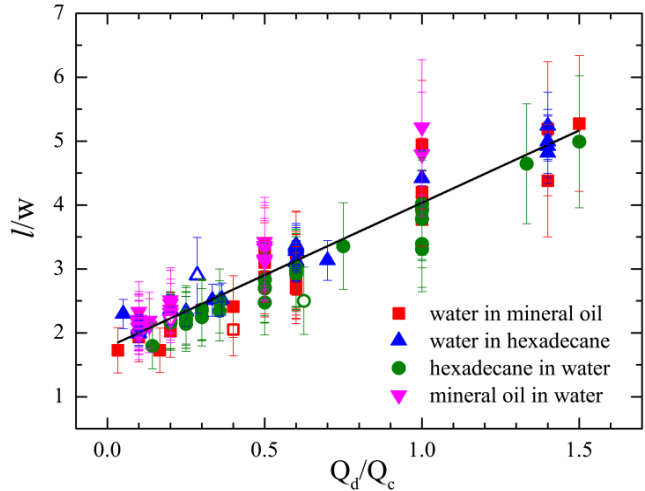

**Fig. 5.** Normalized droplets length as a function of the ratio between the flow rates of the dispersed and continuous phases. Each point corresponds to a different chip. The empty symbols refer to the data presented in Fig. 4. The error bars are mainly due to the uncertainty in the determination of the channels width w. The straight line is a linear fit to all the data.

The chip-to-chip reproducibility has been checked by comparing the lengths of water-in-hexadecane and hexadecane-in-water droplets produced in three different devices working at the same flow conditions. The results are summarized in Table 2. The mean size deduced from analysing more than 50 drops was found to change by about 10% among the chips. This figure can be easily improved by using a better plotter. The continuous phases (hexadecane with 1% w/w

Span 80 and distilled water with 1% v/v Triton X-100, respectively) flow rate was 7 µL/min, while that of the dispersed phases (distilled water and hexadecane, respectively) flow rate was 3 µL/min. The mean size was found to change by less than 10%. Again, we mainly attribute this variations to the cheap cutting plotter used in this work. A better one will significantly improve the chip-to-chip reproducibility.

**Table 2.** Mean lengths of water-in-hexadecane and hexadecane-in-water droplets measured in six different chips

| Water-in-hexadecane | | Hexadecane-in-water | |
|---|---|---|---|
| Microchip | Drop Length (mm) | Microchip | Drop Length (mm) |
| 1 | 1.41 ± 0.02 | 1' | 0.49 ± 0.02 |
| 2 | 1.55 ± 0.03 | 2' | 0.62 ± 0.02 |
| 3 | 1.34 ± 0.03 | 3' | 0.56 ± 0.02 |
| Mean | 1.43 ± 0.11 | Mean | 0.56 ± 0.07 |

High-frequency generation of small volume droplets is useful for biological assays as the individual, picoliter-scale microvessels can contain small numbers of molecules or cells which can nevertheless be at biologically relevant concentrations [9]. In a T-junction generator, the droplet frequency depends also on the inflows of the two immiscible liquids: higher flows imply higher frequencies. This implies that the channels must be able to sustain pressures of a few bars. We then evaluated the maximum pressure above which the PeT channels started to delaminate. We exploited the fact that the flow rate Q in a microfluidic channel depends linearly on the pressure gradient $\Delta P$ along it, i.e. $\Delta P = R \cdot Q$, where R is the hydrodynamic resistance of the channel. In particular, we fabricated straight microchannels having cross sections of about 60 µm by 120 µm and length of 2.5 cm. By using a flow controller (MFCS-Flowell, Fluigent) we applied known, small pressure gradients $\Delta P$ across the channels and measured the resulting flow rate Q of distilled water. The graph in Fig. 6 shows a representative set of data. The straight line is the linear fit to the data. The channel flow resistance R is the slope of this curve, which is equal to 1.87 ± 0.02 mBar min/µL. This value compares reasonably well with 2.5 ± 1 mBar min/µL calculated for a straight channel having a rectangular cross section with width w = 60 µm and height h = 120 µm, a length l=2.5 cm and a water viscosity of 0.9 cP corresponding to 25 °C [32]. The large error mainly reflects the uncertainty in the width w of ~ 10 µm due to irregularities in the cross-section as shown in Figure 1. Afterwards, we connected the channel to a syringe pump (PHD 2000, Harvard Apparatus) and we applied an increasing flow rate in steps of 100 µL/min. The channels were found to leak at a flow rate of about 1200 µL/min corresponding to a $\Delta P \sim 2.3$ bar. In conventional PDMS channels used for the production of droplets at high frequency, the typical pressure used to produce droplet are normally below 1.5 bar [33]. Therefore, the PeT chips can be successfully employed in similar conditions.

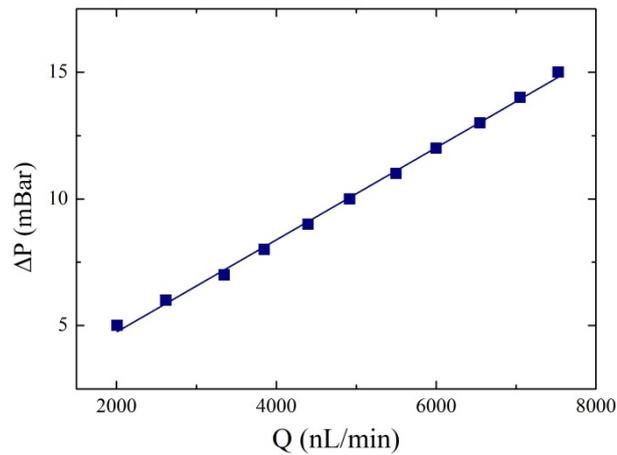

**Fig. 6.** Flow rate Q induced by an increasing pressure gradient ΔP applied across a PeT microchannel.

Actually, we measured the droplet generation of PeT devices at the maximum flow rates reachable with our setup. Droplets of water in hexadecane + 2% w/w Span 80 were produced at a frequency close to 550 Hz with $Q_d$ = 10 μL/min and $Q_c$ = 800 μL/min.. The analysis of the droplets images taken with a fast camera (Phantom VRI v7.3) yielded the histogram shown in Figure 7 which presents a polidispersity of 1.6 %, compatible with those measured at much lower frequencies (a few Hz) of Figure 4. The droplet volume was of about 0.4 nL and, as evident by the snapshot of Figure 7, the PeT device was working in the dripping regime [30,31].

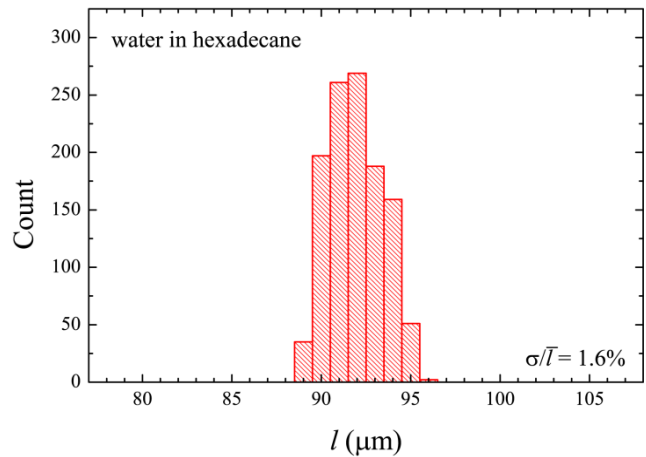

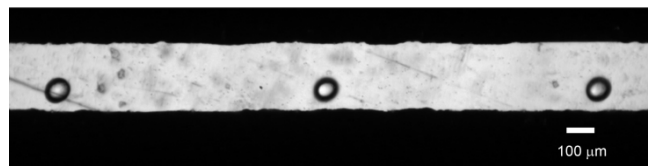

**Fig. 7.** Top: histogram showing the size distribution of water-in-hexadecane droplets produced at a frequency of 550 Hz. Bottom: optical image of the drops flowing in the main channel.

## 4. Conclusion

In conclusion, we have demonstrated that T-junction microfluidic devices constructed in polyester-toner are suitable for generating highly monodisperse W/O and O/W droplets in the same device without any specific surface treatment of the channels walls. This is mainly due to the partial hydrophilicity of the polyester surface together with the addition of suitable surfactants. The fabrication procedure is characterized by being very simple, fast and inexpensive. We estimate that it is possible to complete the whole fabrication in less than 10 minutes, with a cost of ~$ 0.10 per device. This makes the lamination of PeT foils the fastest and cheapest rapid-prototyping technique. Actually, the standard chips in PDMS require the fabrication of a master, typically done by lithography, and a process in oxygen plasma for the sealing of the chip. Furthermore, PDMS is hydrophobic so that the formation of O/W drops needs prior functionalization of the surface walls. On the other hand, very uniform channels of any size can be made by replica molding of PDMS. The poor resistance of PeT and PDMS chips to organic solvents can be overcome using glass chips. The production of W/O drops in these chips requires a functionalization of the surface walls. However, the main limitation of glass chips is the difficulty in making them. There are commercial companies selling them with standard and custom designs at a starting cost of tens of $ per piece. We synthetically compare the PeT chips for drop productions with those made in PDMS and glass in Table 3.

Table 3. Main specs of microfluidic drop generators made in different materials

|  | PeT | PDMS | Glass |
|---|---|---|---|
| Rapid prototyping | Yes | Yes | No |
| Channel size | Minimum height set by PE thickness Minimum width set by plotter blade | Any | Any |
| Channel uniformity | Good (depends on cutting plotter) | Very good | Very good |
| Chip sealing | Lamination | Oxygen plasma | Thermal bonding |
| Equipment required | Very cheap and simple | Standard equipment for microfabrication | Specialized equipment for microfabrication |
| Chemical resistance | No organic solvents | No organic solvents | Very good |
| Water contact angle | 50° | 110° | < 20° |
| Drops combination without surface treatment | W/O and O/W | W/O | O/W |

The characteristics of the PeT chips and the possibility of large-scale production make the polyester-toner technology very attractive for widespread utilization also in droplet-based microfluidics.

## Acknowledgements

E.P. acknowledges a fellowship from Coimbra Group. Grants from European Research Council under the European Community's Seventh Framework Programme (FP7/2007-2013) / ERC Grant Agreement N. 297004 and from the University of Padova (PRAT 2011 'MINET' and PRAT 2009 N. CPDA092517/09) are gratefully acknowledged.

## Authors biographies

**Evandro Piccin** is a professor at the Department of Chemistry, Federal University of Minas Gerais, Brazil. He received his Ph.D. degree in Analytical Chemistry from University of Sao Paulo, in 2008. During his PhD, he has spent one year as exchange student in the Biodesign Institute, Arizona State University, working with microfluidics. His research interests are currently focused on the development of low-cost technologies for fabricating microfluidic devices, integration of microchip electrophoresis and chemical sensors, and chemical analysis in paper-based microfluidic devices.

**Davide Ferraro** received his Bachelor and Master degree in materials science and engineering from Padua University (Italy) in 2007 and 2009, respectively. After that, he obtained his PhD in Materials Science and engineering from Padua University in 2013. Currently he is Post-doc at the Curie Insitut in the group of Jean-Louis Viovy in Paris (France) focusing on droplet-microfluidic platform for genetic screening of single cancer cells.

**Paolo Sartori** received his BS and MS degrees in Physics from Padua University (Italy) in 2007 and 2012, respectively.

He is currently a PhD student at the Department of Physics and Astronomy "G. Galilei" of Padua University (Italy). His research interests include droplet microfluidics, wetting phenomena and active matter.

**Enrico Chiarello** received his BS and MS degrees in Physics from Padua University (Italy) in 2010 and 2013, respectively. He is currently a PhD student at the Department of Physics and Astronomy "G. Galilei" of Padua University (Italy). His research interests include wetting phenomena and active matter.

**Matteo Pierno** obtained the MS degree in Physics at the University of Milan (Italy) in 1999. Then he received the PhD in Engineering in 2004 at the Polytechnic of Milan (Italy) dealing with fluorinated sols and films. From 2004 to 2006 he has been Marie Curie Post-Doc in Montpellier (France) within the frame of the Research and Training Network on the Dynamical Arrested State of Soft Matter and Colloids. Starting from 2007 he is assistant professor at the Department of Physics and Astronomy of the University of Padua (Italy). Research interests are focused on soft-matter, liquids at interfaces and nanotribology.

**Giampaolo Mistura** received his MS degree in physics from Padua University (Italy) in 1986. He obtained his PhD degree in Physics from Penn State University (USA) in 1993, working on the wetting properties of cryogenic fluids. He spent one year as a post doc at the University of Konstanz (Germany) and one at the High Magnetic Field Laboratory in Grenoble (France). Since 1996 he is at the Physics Department of Padua University (Italy) where he is currently associate professor. His main research activities include the study of interfacial phenomena (wetting, adsorption and nanofriction) and microfluidics.